\documentstyle[12pt,a4wide]{article}
\newlength{\absize}
\renewcommand{\arraystretch}{2.0}

\begin{document}
\thispagestyle{empty}
\pagestyle{empty}
\renewcommand{\thefootnote}{\fnsymbol{footnote}}
\newcommand{\starttext}{\newpage\normalsize
  \pagestyle{plain}
  \setlength{\baselineskip}{4ex}\par
  \setcounter{footnote}{0}
  \renewcommand{\thefootnote}{\arabic{footnote}}}
\newcommand{\preprint}[1]{%
  \begin{flushright}
    \setlength{\baselineskip}{3ex} #1
  \end{flushright}}
\renewcommand{\title}[1]{%
  \begin{center}
    \LARGE #1
  \end{center}\par}
\renewcommand{\author}[1]{%
  \vspace{2ex}
  {\Large
   \begin{center}
     \setlength{\baselineskip}{3ex} #1 \par
   \end{center}}}
\renewcommand{\thanks}[1]{\footnote{#1}}
\renewcommand{\abstract}[1]{%
  \vspace{2ex}
  \normalsize
  \begin{center}
    \centerline{\bf Abstract}\par
    \vspace{2ex}
    \parbox{\absize}{#1\setlength{\baselineskip}{2.5ex}\par}
  \end{center}}

\setlength{\parindent}{3em}
\setlength{\footnotesep}{.6\baselineskip}
\newcommand{\myfoot}[1]{%
  \footnote{\setlength{\baselineskip}{.75\baselineskip}#1}}
\renewcommand{\thepage}{\arabic{page}}
\setcounter{bottomnumber}{2}
\setcounter{topnumber}{3}
\setcounter{totalnumber}{4}
\newcommand{\figsize}{}
\renewcommand{\bottomfraction}{1}
\renewcommand{\topfraction}{1}
\renewcommand{\textfraction}{0}
%%%%%%%%%%%%%%%%%%%%%%%%%%%%%%%%%%%%%%%%%%%%%%%%%%%%%%%%%%%%%%%%%%%%%%%%
%
% More commands and definitions:
%
\newcommand{\beq}{\begin{equation}}
\newcommand{\eeq}{\end{equation}}
\newcommand{\beqa}{\begin{eqnarray}}
\newcommand{\eeqa}{\end{eqnarray}}
\newcommand{\nn}{\nonumber}

\newcommand{\dd}{{\rm d}}
\newcommand{\mH}{m_{\rm H}}
\newcommand{\mgluino}{m_{\tilde g}}
\newcommand{\msquark}{m_{\tilde q}}
\newcommand{\gluino}{\tilde g}
\newcommand{\squark}{\tilde q}
\newcommand{\alphas}{\alpha_{\rm s}}
\newcommand{\mZ}{m_Z}
\newcommand{\Cplus}{C^+_{\tilde u}}
\newcommand{\Cminus}{C^-_{\tilde u}}
\newcommand{\Ctilu}{C_{\tilde u}}
\newcommand{\Ctilq}{C_{\tilde q}}
\newcommand{\thW}{\theta_{\rm W}}
\newcommand{\thu}{\theta_{\tilde u}}
\newcommand{\thd}{\theta_{\tilde d}}
\newcommand{\GeV}{\mbox{{\rm GeV}}}
\newcommand{\gs}{g_{\rm s}}
\renewcommand{\Re}{{\rm Re}}
\renewcommand{\Im}{{\rm Im}}

%%%%%%%%%%%%%%%%%%%%%%%%%%%%%%%%%%%%%%%%%%%%%%%%%%%%%%%%%%%%%%%%%%%%%%%%
%
% More commands and definitions by Bjarte:
%
\newcommand{\bas}{\begin{eqnarray*}}
\newcommand{\eas}{\end{eqnarray*}}
\newcommand{\bg}{\begin{equation}}
\newcommand{\ed}{\end{equation}}
\newcommand{\ba}{\begin{eqnarray}}
\newcommand{\ea}{\end{eqnarray}}
\newcommand{\hc}{\mbox{{\rm h.c.}}}
\newcommand{\Tr}{\mbox{Tr}}
\newcommand{\what}[3]{\settowidth{\ltT}{\Dis{#3}}
\makebox[\ltT]{$\rule{#2\mmh}{0mm}
\widehat{\makebox[#1\mm]{\Dis{#3\rule{#2\mm}{0mm}}}}$}}
\newcommand{\hl}{h_{{ l}}}
\newcommand{\Tsp}{\mbox{\scriptsize T}}
\newcommand{\hdq}{h_{{ d}}}
\newcommand{\huq}{h_{{ u}}}
\newcommand{\thth}{\theta\theta}
\newcommand{\othth}{\overline{\theta}\,\overline{\theta}}
\newcommand{\aeep}{\tilde{{ e}}}
\newcommand{\mW}{m_{{ W}}}
\newcommand{\me}{m_{{ e}}}
\newcommand{\Ae}{A_{{ e}}}
\newcommand{\Md}{m_{{ d}}}
\newcommand{\Ad}{A_{{ d}}}
\newcommand{\sdq}{\tilde{{ d}}}
\newcommand{\Mu}{m_{{ u}}}
\newcommand{\Au}{A_{{ u}}}
\newcommand{\suq}{\tilde{{ u}}}
\newcommand{\suqa}{\tilde{{ u}}_{1}}
\newcommand{\suqb}{\tilde{{ u}}_{2}}
\newcommand{\sdqa}{\tilde{{ d}}_{1}}
\newcommand{\sdqb}{\tilde{{ d}}_{2}}
\newcommand{\squ}{\wtilde{2}{0}{q}}
\newcommand{\glip}{\wtilde{3}{0.2}{g}}
\newcommand{\T}{\mbox{T}}

\newcommand{\msd}{m_{\tilde{d}}}
\newcommand{\msu}{m_{\tilde{u}}}
\newcommand{\MMse}{\wtilde{3}{0.8}{M}_{\hspace*{-1.2mm}E}}
\newcommand{\Msle}{\wtilde{3}{0.2}{m}_{\!E}}
\newcommand{\MsQU}{\wtilde{3}{0.8}{M}_{\hspace*{-1mm}U}}
\newcommand{\MsQT}{\wtilde{3}{0.8}{M}_{\hspace*{-1mm}T}}
\newcommand{\Msq}{m_{\tilde{q}}}
\newcommand{\Msu}{\wtilde{3}{0.2}{m}_{U}}
\newcommand{\Msd}{\wtilde{3}{0.2}{m}_{\!D}}
\newcommand{\MsB}{\wtilde{3}{0.2}{m}_{B}}
\newcommand{\MsT}{\wtilde{3}{0.2}{m}_{T}}
\newcommand{\jots}[1]{\setlength{\jot}{#1 mm}}
\newcommand{\Dis}[1]{$\displaystyle #1$}
\newcommand{\eps}{\epsilon}
\newcommand{\wtilde}[3]{\settowidth{\ltT}{\Dis{#3}}
\makebox[\ltT]{$\rule{#2\mmh}{0mm}
\widetilde{\makebox[#1\mm]{\Dis{#3\rule{#2\mm}{0mm}}}}$}}
\newcommand{\MHe}{M_{\!H_{1}}}
\newcommand{\MHt}{M_{\!H_{2}}}
\newcommand{\mlN}{m_{\Disp{\tilde{{\scriptstyle \lambda}}}}}
\newcommand{\MNS}{m_{\Disp{\tilde{{\scriptstyle \Lambda}}}}}
\newcommand{\ola}{\overline{\lambda}}
\newcommand{\oLa}{\overline{\Lambda}}
\newcommand{\mgli}{m_{\tilde{g}}}
\newcommand{\gli}{\psi_{g}}
\newcommand{\ogli}{\overline{\psi}_{g}}
\newcommand{\Disp}[1]{{\displaystyle #1}}
\newcommand{\tu}{\theta_{\tilde{u}}}
\newcommand{\td}{\theta_{\tilde{d}}}
\newcommand{\Msqa}[1]{m_{\tilde{q}_{#1}}}
\newcommand{\tq}{\theta_{\tilde{q}}}
\newcommand{\tqu}{\tq}

\newlength{\ltT}
\newlength{\mmh}
\setlength{\mmh}{0.5mm}
\newlength{\mm}
\setlength{\mm}{1mm}

\newcommand{\Qhat}{\what{3}{0.7}{Q}}
\newcommand{\Glue}{\wtilde{3}{0.8}{G}}
%%%%%%%%%%%%%%%%%%%%%%%%%%%%%%%%%%%%%%%%%%%%%%%%%%%%%%%%%%%%%%%%%%%%%%%%%%
\def\draftmode#1{\def\draftdate{
\number\day-\number\month-\number\year}
\headline={\hfil#1:\ \draftdate}}

\def\draftdate{
\number\day-\number\month-\number\year}

\def\Draft#1{{\hfil#1:\ \draftdate}}

\def\Month{\ifcase\month\or
January\or February\or March\or April\or May\or June\or
July\or August\or September\or October\or November\or December\fi}
\def\slash#1{#1 \hskip -0.5em /}
\def\qslash{\slash q}
\def\kslash{\slash k}
\def\ieps{i\epsilon}
\def\Order{{\cal O}}
%%%%%%%%%%%%%%%%%%%%%%%%%%%%%%%%%%%%%%%%%%%%%%%%%%%%%%%%%%%%%%%%%%%%%%%%
%
% Here we go:

\preprint{University of Bergen, Department of Physics \\
Scientific/Technical Report No.\ 1994-10 \\ ISSN~0803-2696 \\
\Month, \the\year}

\vfill
\title{Gluino Production in Electron-Positron Annihilation}

\vfill
\author{Bjarte Kileng \\ Per Osland \\\hfil\\
        Department of Physics\thanks{electronic mail addresses:
                {\{kileng,osland\}@vsfys1.fi.uib.no}}\\
        University of Bergen \\ All\'egt.~55, N-5007 Bergen, Norway }
\date{}

%%%%%%%%%%%%%%%%%%%%%%%%%%%%%%%%%%%%%%%%%%%%%%%%%%%%%%%%%%%%%%%%%%%%%%%%
\vfill
\abstract{We discuss the pair production of gluinos in electron-positron
annihilation at LEP, in a model with soft supersymmetry breaking,
allowing for mixing between the squarks.
In much of the parameter space of the Minimal Supersymmetric Model (MSSM)
the cross section corresponds
to a $Z$ branching ratio above $10^{-5}$, even up to $10^{-4}$.
A non-observation of gluinos at this level
restricts the allowed MSSM parameter space.
In particular, it leads to lower bounds on the soft mass
parameters in the squark sector.
}

\vfill

%%%%%%%%%%%%%%%%%%%%%%%%%%%%%%%%%%%%%%%%%%%%%%%%%%%%%%%%%%%%%%%%%%%%%%%%
\starttext
%%%%%%%%%%%%%%%%%%%%%%%%%%%%%%%%%%%%%%%%%%%%%%%%%%%%%%%%%%%%%%%%%%%%%%%%
\section{Introduction}
\label{sec:intro}
\setcounter{equation}{0}
%\pagestyle{myheadings}
%\markboth{\Draft{Draft}}{\Draft{Draft}}

Recent searches for gluinos by the CDF Collaboration have
established a lower mass bound of the order of
140~GeV/${\rm c}^2$ \cite{CDF}.
This bound depends on the assumed decay mode of the gluino,
it is valid for the case of direct decay to the lightest
supersymmetric particle, $\gluino\to q \bar q \tilde\chi$.
The analysis is insensitive to light gluinos,
$\mgluino\le\Order(40\ \GeV)$.
However, various other experiments, in particular those at the CERN SPS
\cite{UA1,UA2}
exclude most of the region below
40~GeV, except for a narrow range around 3--5~GeV/${\rm c}^2$ \cite{PDG}.

The existence of this low-mass gluino window has recently been
pointed out
\cite{UA1},
and it is even argued that data on $\alphas(\mZ)$ favour
a light gluino
\cite{Jezabek,Clavelli,Ellis}.
(See also ref.\ \cite{Hebbeker}, however.)
Some of its further consequences are explored in ref.\ \cite{Campos}.

The importance of searching for light gluinos has long been stressed
\cite{Altar}.
Clearly, if the gluino is very light, it should be produced at LEP,
either by radiation in pairs off a quark \cite{Farrar,Campbell1},
or in pairs via
the triangle diagram \cite{Nelson,Kane,Campbell2}.
In the former case, the final four-jet state would be rather
hard to isolate, because of the QCD background
\cite{Hultqvist}.
For the latter mechanism, the cross section was at low energies
(ref.~\cite{Nelson}, photon exchange) found to
depend very much on the mass splitting between the squarks,
being in general rather small.
A similar analysis has been performed for the $Z$ decay
\cite{Kane,Campbell2}, and the cross section was found to depend
sensitively on the mass splitting between the top and bottom quarks.
Because the previous analyses are limited to low
top-quark masses, and in order to also study the effects of
chiral mixing, we find it important to present a new analysis
of the gluino pair production cross section.

The notation to be used is in part given by the MSSM Lagrangian
density
\begin{equation} {\cal L} = {\cal L}_{\mbox{{\scriptsize SU(3)}}}
 + {\cal L}_{\mbox{{\scriptsize SU(2)$\times$U(1)}}}
 + {\cal L}_{\mbox{{\scriptsize Soft}}} \hspace*{1mm},
\label{EQU:Lagrangeone}
\end{equation}
with the SU(3) part given by (subscripts ``s" for ``strong")
\begin{equation} {\cal L}_{\mbox{{\scriptsize SU(3)}}} =
   \left\{ \frac{1}{8\gs^2}
   \Tr\left[ W_{\rm s}^{\alpha}W_{\rm s\alpha} \right]_{\thth}
 + \hc \right\}
 + \left[ \Qhat^{L\dagger}_{\mbox{\scriptsize SU(3)}}e^{2\gs
   V_{\rm s}}\Qhat^{L}_{\mbox{\scriptsize SU(3)}}
  + \Qhat^{R}_{\mbox{\scriptsize SU(3)}}e^{-2\gs V_{\rm s}}
    \Qhat^{R\dagger}_{\mbox{\scriptsize SU(3)}}
    \right]_{\thth\othth} \hspace*{1mm}, \end{equation}
and the SU(2)$\times$U(1) part by
\jots0
\begin{eqnarray} {\cal L}_{\mbox{{\scriptsize SU(2)}}\times
    \mbox{{\scriptsize U(1)}}} & = & \Biggl\{ \biggl[
    \frac{1}{8g^{2}}\Tr\left[ W^{\alpha}W_{\alpha} \right]
    + \frac{1}{4}w^{\alpha}w_{\alpha}
    - \mu \what{3}{0.7}{H}^{\Tsp}_{1} \eps \what{3}{0.7}{H}_{2}
    \nonumber \\
& & \hspace*{15mm} \mbox{}
    + \hl\what{3}{0.7}{L}^{\Tsp}\eps \what{3}{0.7}{H}_{1}
      \what{3}{0.7}{E}^{R}
    + \hdq\Qhat^{\Tsp}\eps\what{3}{0.7}{H}_{1}
      \what{3}{0.7}{D}^{R}
    + \huq\Qhat^{\Tsp}\eps\what{3}{0.7}{H}_{2}
      \what{3}{0.2}{U}^{R}
 \biggr]_{\thth}
 + \hc  \Biggr\} \nonumber \\[5mm]
& & + \Biggl[ \what{3}{0.7}{L}^{\dagger}e^{\left( 2gV - g'v \right)}
      \what{3}{0.7}{L}
    + \what{3}{0.7}{E}^{R\dagger}e^{2g'v}\what{3}{0.7}{E}^{R}
    + \Qhat^{\dagger}e^{(2gV+g'v/3)}\Qhat
    + \what{3}{0.2}{U}^{R\dagger}e^{-4g'v/3}\what{3}{0.2}{U}^{R}
      \nonumber \\
& & \hspace*{15mm} \mbox{}
    + \what{3}{0.7}{D}^{R\dagger}e^{2 g'v/3}\what{3}{0.7}{D}^{R}
    + \what{3}{0.7}{H}^{\dagger}_{1}e^{ (2gV-g'v) }\what{3}{0.7}{H}_{1}
    + \what{3}{0.7}{H}^{\dagger}_{2}e^{ (2gV+g'v) }\what{3}{0.7}{H}_{2}
      \Biggr]_{\thth\othth} \label{hhcoup} \hspace*{1mm}.
\end{eqnarray}
Here, ``hats" refer to superfields.
The gauge--invariant (soft) supersymmetry breaking part is given
in terms of component fields as
\begin{eqnarray}
\label{EQU:Lagrangefour}
{\cal L}_{\mbox{{\scriptsize Soft}}}
& = & \Biggl\{
      \beta^{Hh} H_{1}^{\Tsp} \eps H_{2}
    + \frac{g\me\Ae}{\sqrt{2}\;\mW\cos\beta}L^{\Tsp}\eps H_{1}\aeep^{R}
      \nonumber \\
& & \hspace*{15mm} \mbox{} + \frac{g\Md\Ad}{\sqrt{2}\;\mW\cos\beta}
    Q^{\Tsp}\eps H_{1}\sdq^{R}
    - \frac{g\Mu\Au}{\sqrt{2}\;\mW\sin\beta}Q^{\Tsp}\eps H_{2}\suq^{R}
    + \hc \Biggr\} \nonumber \\[7mm]
& & -  \MMse^{2}L^{\dagger}L - \Msle^{2}\aeep^{R\dagger}\aeep^{R}
    - \MsQU^{2}Q^{\dagger}Q - \Msu^{2}\suq^{R\dagger}\suq^{R}
    - \Msd^{2}\sdq^{R\dagger}\sdq^{R} \nonumber \\[3mm]
& & - \MHe^{2} H_{1}^{\dagger}H_{1} - \MHt^{2} H_{2}^{\dagger}H_{2}
    + \frac{\mlN}{2}\left(\lambda\lambda +\ola\;\ola\right)
    + \frac{\MNS}{2} \sum_{I=1}^{3} \left(\Lambda_{I}\Lambda_{I}
    + \oLa_{I}\oLa_{I}\right) \nonumber \\
& & + \frac{\mgli}{2} \sum_{a=1}^{8} \left({\gli}_{a}{\gli}_{a}
    + {\ogli}_{a}{\ogli}_{a}\right)   \hspace*{1mm}.
\end{eqnarray}
Subscripts $u$ (or $U$) and $d$ (or $D$) refer generically to up and
down-type quarks.  We shall mostly focus on the contributions
from the third generation.  Thus, these symbols will actually
often refer to top and bottom quarks.
Spinors are here expressed in two-component Weyl notation,
since the chiral mixing acts at this level.
The notation is further explained in ref.~\cite{Kileng} and references
quoted there.

The gluino mass is given explicitly by $\mgluino$, whereas squark
masses depend not only on the explicit mass parameters $\MsQU$,
$\Msu$ and $\Msd$, but also on $m_u$, $m_d$, $m_Z$, $m_W$,
$A_u$, $A_d$, $\mu$ and $\beta$.
For each flavour, there are two squarks, whose masses are given
in terms of a similar parameterization in ref.~\cite{Brignole}.
(See also ref.~\cite{Kileng}.)
In the limit of no mixing, i.e., with $\mu=0$, and
$A_d=A_u=0$, the masses of the squarks associated with
left- (L) and right- (R) chiral quarks are given by
\jots2
\beqa
\label{EQU:massnomix}
m^2_{\tilde u\,{\rm L}}&=& m_u^2+\MsQU^2
-\left({1\over6}m_Z^2-{2\over3}m_W^2 \right) \cos(2\beta), \nn \\
m^2_{\tilde u\,{\rm R}}&=& m_u^2+\MsQU^2
+\left({2\over3}m_Z^2-{2\over3}m_W^2\right) \cos(2\beta), \nn \\
m^2_{\tilde d\,{\rm L}}&=& m_d^2+\MsQU^2
-\left({1\over6}m_Z^2+{1\over3}m_W^2 \right) \cos(2\beta), \nn \\
m^2_{\tilde d\,{\rm R}}&=& m_d^2+\MsQU^2
-\left({1\over3}m_Z^2-{1\over3}m_W^2\right) \cos(2\beta).
\eeqa
We shall however consider the case of mixing, for which
the mass formulas are more complicated \cite{Kileng,Brignole}.

It should be noted that the above Lagrangian represents a model
which is different from the recently considered
``constrained" models based on Grand Unification and supergravity
\cite{RGR,KaneCMMSM}.
In particular, the gluino mass is here not tied to the other
gaugino masses.
The model is ``minimal" in the sense that it has only two Higgs
doublets, the soft mass terms are however ``non-minimal".

%%%%%%%%%%%%%%%%%%%%%%%%%%%%%%%%%%%%%%%%%%%%%%%%%%%%%%%%%%%%%%%%%%%%%%%%%%%
\section{The $Z\gluino\gluino$ amplitudes}
\setcounter{equation}{0}
\label{sec:ampl}
In the decay of the $Z$, or more generally in electron-positron
annihilation, the pair production of gluinos can proceed via the two generic
diagrams $(a)$ and $(b)$ of figure~1, where the internal lines
of the triangles are quarks and squarks.
Allowing for mixing between the squarks associated with
the left- and right-chiral quark superfields, we find
the Feynman rules for the vertices as given in figure~2.

We shall write the amplitude for
\beq
e^+e^- \to \gluino\gluino
\eeq
as
\beq
\label{EQU:amplitude}
{\cal M} = {L}^{\mu}iD_{{\rm F}\mu\nu}{\Glue}^{\nu}\,\delta_{ab},
\eeq
where the lepton current is given as
\beq
{L}^{\mu} = \overline{v}(p_{2})
\left\{ \frac{-ig\gamma^{\mu}}{2\cos\thW}\left( g_{V} - g_{A}\gamma_{5} \right)
\right\} u(p_{1}),
\eeq
and the gluino current $\Glue^\mu$ will consist of
a sum over contributions from
different diagrams to be discussed presently.
Furthermore, $iD_{{\rm F}\mu\nu}$ is the $Z$ propagator, and
$\delta_{ab}$ is a Kronecker delta in the gluino colour indices.
For each quark flavour, there are two uncrossed and two crossed diagrams of
type~$(a)$.  If we label them by the quark and squark propagators of the
triangle, then we can write the terms involving $u$-quarks as
\beqa
\label{EQU:Mpruncrossed}
\Glue^\mu_{uu1}
&=& -N_u\, \bar u(k_2)(\Cplus-\Cminus\gamma_5)
T^\mu_{uu1}(k_1,k_2)(\Cplus+\Cminus\gamma_5)\, C^{-1}
\bar u^{\rm T}(k_1), \nn \\
\Glue^\mu_{uu2}
&=& -N_u\, \bar u(k_2)(\Cminus+\Cplus\gamma_5)
T^\mu_{uu2}(k_1,k_2)(\Cminus-\Cplus\gamma_5)\, C^{-1}
\bar u^{\rm T}(k_1).
\eeqa
Here, $C$ denotes the charge conjugation matrix and T transposition.
Since the gluino is a Majorana fermion, the currents contain the factor
$C^{-1} \bar u^{\rm T}(k_1)$ rather than the $v(k_1)$ associated with
Dirac fermions, but one could alternatively have used antiparticle
spinors of opposite spins \cite{JonesLS}.
However, this is less convenient in dealing with the interference
terms between uncrossed and crossed diagrams.
Furthermore, the subscripts 1 and 2 refer to
the mass eigenstates of the squarks.
The quark-squark-gluino couplings depend on the chiral mixing
(see figure~2), and
are proportional to the coefficients
\beq
C^{\pm}_{\tilde u}=\cos\thu \pm\sin\thu.
\eeq
Furthermore,
\beq
N_u={g\,\gs^2\over16(2\pi)^4\cos\thW}.
\eeq
Here, $g$ and $\gs$ are the $SU(2)$ and $QCD$ coupling constants.
For photon exchange, the corresponding factor is
\beq
N_u=\frac{e \gs^{2}}{6(2\pi)^{4}}.
\eeq
The triangle integral associated with this diagram $(a)$ is given by
\beqa
\label{EQU:trianglea}
T^\mu_{uui}(k_1,k_2)
&=&{\int}\dd^4 q {\qslash+\kslash_1+m_u\over (q+k_1)^2-m_u^2+\ieps}
\,\gamma^\mu(g_V^u-g_A^u\gamma_5) \nn \\
&{}&\times{\qslash-\kslash_2+m_u\over (q-k_2)^2-m_u^2+\ieps}\,
{1\over q^2-m_i^2+\ieps},
\eeqa
with
\beq
g_V^u=1-{8\over3}\sin^2\thW, \qquad g_A^u=1.
\eeq

For each quark flavour, there are also four uncrossed
and four crossed diagrams of type $(b)$.
The gluino currents corresponding to the uncrossed diagrams involving
the $u$-quarks can be written as
\beqa
\label{EQU:Muncrossed}
{\Glue}_{11u}^{\mu}
&=& -N_{11u} \bar u(k_2)(\Cplus-\Cminus\gamma_5)T^\mu_{11u}(k_1,k_2)
(\Cplus+\Cminus\gamma_5)\, C^{-1}
\bar u^{\rm T}(k_1), \nn \\
{\Glue}_{22u}^{\mu}
&=& -N_{22u} \bar u(k_2)(\Cminus+\Cplus\gamma_5)
T^\mu_{22u}(k_1,k_2)(\Cminus-\Cplus\gamma_5)\, C^{-1}
\bar u^{\rm T}(k_1), \nn \\
{\Glue}_{12u}^{\mu}
&=& -N_{12u} \bar u(k_2)(\Cplus-\Cminus\gamma_5)
T^\mu_{12u}(k_1,k_2)(\Cminus-\Cplus\gamma_5)\, C^{-1}
\bar u^{\rm T}(k_1), \nn \\
{\Glue}_{21u}^{\mu}
&=&-N_{21u} \bar u(k_2)(\Cminus+\Cplus\gamma_5)
T^\mu_{21u}(k_1,k_2)(\Cplus+\Cminus\gamma_5)\, C^{-1}
\bar u^{\rm T}(k_1),
\eeqa
with over-all factors
\beqa
N_{11u}
&=&
\frac{g\,\gs^2}{8(2\pi)^4\cos\thW}
\left(\frac{4}{3}\sin^{2}\thW - \cos^{2}\tu\right), \nn \\
N_{22u}
&=&
\frac{g\,\gs^2}{8(2\pi)^4\cos\thW}
\left(\frac{4}{3}\sin^{2}\thW - \sin^{2}\tu \right), \nn \\
N_{12u}
&=&N_{21u}
=\frac{g\,\gs^2}{16(2\pi)^4\cos\thW}\,\sin(2\tu) \; .
\eeqa
For photon exchange, the corresponding factors are
\beqa
N_{11u}
&=&
N_{22u} = -\frac{e\, g_{\rm s}^{2}}{6(2\pi)^{4}}, \nonumber \\
N_{12u}
&=&N_{21u} = 0.
\eeqa
The triangle integral associated with this diagram $(b)$ is
\beqa
\label{EQU:triangleb}
T^\mu_{iju}(k_1,k_2)
&=&\int\dd^4 q \frac{\qslash+m_u}{q^2-m_u^2+\ieps}
(2q^\mu+k_1^\mu-k_2^\mu) \nn \\
&{}&\times{1\over (q+k_1)^2-m_i^2+\ieps}\,
{1\over (q-k_2)^2-m_j^2+\ieps}.
\eeqa

We need to also discuss the structure of the triangle integrals
in terms of Dirac matrices.
It is convenient to expand the first one, eq.~(\ref{EQU:trianglea}),
in terms of ``even" ($E$) and ``odd" ($O$) scalar integrals as
\beqa
\label{EQU:triangleadecomp}
T^\mu_{uui}(k_1,k_2)
&=& E^{a}_{uui\,\alpha} \gamma^{\alpha}\gamma^{\mu}(g_V^u-g_A^u\gamma_5)
    + E^{b}_{uui\,\alpha} \gamma^{\mu}\gamma^{\alpha} (g_V^u+g_A^u\gamma_5)
\nn \\
& & + O^{a}_{uui} \gamma^{\mu}(g_V^u-g_A^u\gamma_5)
    + O^{b}_{uui\,\alpha\beta} \gamma^{\alpha}\gamma^{\mu}\gamma^{\beta}
(g_V^u+g_A^u\gamma_5),
\label{treq}
\eeqa
and the other one, eq.~(\ref{EQU:triangleb}) as
\beq
\label{EQU:trianglebdecomp}
T^\mu_{iju}(k_1,k_2)
=E_{iju}^{\mu} + O_{iju\,\alpha}^{\mu}\gamma^{\alpha} . \label{treqb}
\eeq
These integrals are discussed in Appendix A.

The gluino current of eq.~(\ref{EQU:amplitude}) can now be written as
\beq
\label{EQU:gluesum}
\Glue^\mu = \sum_{\mbox{\scriptsize flavours}} \Glue^\mu_q,
\eeq
with the $u$-quark contribution
\beqa
\label{EQU:glueu}
\Glue^\mu_u
&=&
\left(\Glue^\mu_{uu1} + \Glue^\mu_{uu2}\right)
+ \left({\Glue}_{11u}^{\mu} + {\Glue}_{22u}^{\mu}
+       {\Glue}_{12u}^{\mu} + {\Glue}_{21u}^{\mu}\right) \nn \\
& & + \mbox{crossed terms}
\eeqa

For each diagram there is a crossed diagram, whose amplitude is obtained by
interchanging the gluino momenta, $k_1\leftrightarrow k_2$, and reversing the
over-all sign.
Thus, the first terms of the amplitudes corresponding to
the crossed diagrams are obtained from
eqs.~(\ref{EQU:Mpruncrossed}) and (\ref{EQU:Muncrossed}) as
\beqa
\Glue_{uu1}^{\mu\rm (cr)}
&=& N_u\, \bar u(k_1)(\Cplus-\Cminus\gamma_5)
T^\mu_{uu1}(k_2,k_1)(\Cplus+\Cminus\gamma_5)\, C^{-1}
\bar u^{\rm T}(k_2), \nn \\
\Glue_{11u}^{\mu\rm (cr)}
&=& N_{11u} \bar u(k_1)(\Cplus-\Cminus\gamma_5)
T^\mu_{11u}(k_2,k_1)(\Cplus+\Cminus\gamma_5)\, C^{-1}
\bar u^{\rm T}(k_2).
\eeqa

Furthermore, there are $4+8$ amplitudes involving the $d$-quark,
with chiral mixing given by
\beq
C^{\pm}_{\tilde d}=\cos\thd \pm\sin\thd,
\eeq
over-all factors,
\beqa
N_d
&=&
N_u, \nn \\
N_{11d}
&=&
- \frac{g\,\gs^2}{8(2\pi)^4\cos\thW}
\left(\frac{2}{3}\sin^{2}\thW - \cos^{2}\td\right) \nn \\
N_{22d}
&=&
- \frac{g\,\gs^2}{8(2\pi)^4\cos\thW}
\left(\frac{2}{3}\sin^{2}\thW - \sin^{2}\td \right) \nn \\
N_{12d}
&=&N_{21d}
= -\frac{g\,\gs^2}{16(2\pi)^4\cos\thW}\,\sin(2\td),
\eeqa
and
\beq
g_V^d=-1+{4\over3}\sin^2\thW, \qquad g_A^d=-1.
\eeq

\medskip
\noindent{\bf The no-mixing limit} \\
For comparison, we quote also the simple forms obtained for
the amplitudes (\ref{EQU:Mpruncrossed}) and (\ref{EQU:Muncrossed})
in the limit of no mixing (nm) between the squarks,
\beqa
\Glue_{uu1}^{\mu\rm (nm)}
&=& -N_u\, \bar u(k_2)(1-\gamma_5)
T^\mu_{uu1}(k_1,k_2)(1+\gamma_5)\, C^{-1}
\bar u^{\rm T}(k_1), \nn \\
\Glue_{uu2}^{\mu\rm (nm)}
&=& -N_u\, \bar u(k_2)(1+\gamma_5)
T^\mu_{uu2}(k_1,k_2)(1-\gamma_5)\, C^{-1}
\bar u^{\rm T}(k_1),
\eeqa
and
\beqa
\label{EQU:Mnomix}
\Glue_{11u}^{\mu\rm (nm)}
&=& -N_{11u}^{\rm (nm)} \bar u(k_2)(1-\gamma_5)
T^\mu_{11u}(k_1,k_2)(1+\gamma_5)\, C^{-1}
\bar u^{\rm T}(k_1), \nn \\
\Glue_{22u}^{\mu\rm (nm)}
&=& -N_{22u}^{\rm (nm)} \bar u(k_2)(1+\gamma_5)
T^\mu_{22u}(k_1,k_2)(1-\gamma_5)\, C^{-1}
\bar u^{\rm T}(k_1), \nn \\
\Glue_{12u}^{\mu\rm (nm)}
&=& \Glue_{21u}^{\mu\rm (nm)} = 0,
\eeqa
with
\beqa
N_{11u}^{\rm (nm)}
&=&
-{g\,\gs^2\over8(2\pi)^4\cos\thW}
\left(1-{4\over3}\sin^2\thW\right), \nn \\
N_{22u}^{\rm (nm)}
&=&
{g\,\gs^2\over8(2\pi)^4\cos\thW}
\,{4\over3}\sin^2\thW.
\eeqa
Indices 1 and 2 will then refer to the squarks associated
with the left- and right-chiral quarks.
Their masses are given by eq.~(\ref{EQU:massnomix}).
In the presence of mixing, however, indices 1 and 2 will refer
to the heavier and lighter of the two squarks, respectively.

%%%%%%%%%%%%%%%%%%%%%%%%%%%%%%%%%%%%%%%%%%%%%%%%%%%%%%%%%%%%%%%%%%%%%%%%%%%
\section{The Gluino Current}
\setcounter{equation}{0}
\label{:glucurr}
The gluino current (\ref{EQU:glueu})
can be written as a sum of pairs of terms, corresponding to the
uncrossed and crossed diagrams.
Furthermore, there are 8 amplitudes with two, and 4 with one
internal squark line,
a total of twelve diagrams for each quark flavour.
For the $u$-quark loops we have
\beqa
\label{EQU:eqtwosix}
\Glue^\mu_u
&=&
\sum_i \bigl[\bar u(k_2) M^\mu_{uui}(k_1,k_2) C^{-1} \bar u^{\rm T}(k_1)
+\bar u(k_1) M_{uui}^{\mu\rm (cr)}(k_1,k_2) C^{-1} \bar u^{\rm T}(k_2) \bigr]
\nn \\
&{}&+\sum_{i,j}\bigl[\bar u(k_2) M^\mu_{iju}(k_1,k_2) C^{-1} \bar u^{\rm
T}(k_1)
+\bar u(k_1) M_{iju}^{\mu\rm (cr)}(k_1,k_2)C^{-1} \bar u^{\rm T}(k_2) \bigr]
\nn \\
&=&
\sum_i \bigl\{\bar u(k_2) \bigl[
M^\mu_{uui}(k_1,k_2) -C^{-1} M_{uui}^{\mu\rm (cr)T}(k_1,k_2)C\bigr]
C^{-1} \bar u^{\rm T}(k_1) \bigr\} \nn \\
&{}&+\sum_{i,j}\bigl\{\bar u(k_2) \bigl[
M^\mu_{iju}(k_1,k_2) -C^{-1} M_{iju}^{\mu\rm (cr)T}(k_1,k_2)C\bigr]
C^{-1} \bar u^{\rm T}(k_1) \bigr\},
\eeqa
where in the last step we have transposed the crossed terms, using
$C^{-1{\rm T}}=-C^{-1}$.
It follows from eqs.~(\ref{EQU:Mpruncrossed}) and (\ref{EQU:Muncrossed}) that
\beqa
\label{EQU:threetwo}
M_{uu1}^\mu
&=& -N_u\, (\Cplus-\Cminus\gamma_5)
T^\mu_{uu1}(k_1,k_2)(\Cplus+\Cminus\gamma_5), \nn \\
M_{uu2}^\mu
&=& -N_u\, (\Cminus+\Cplus\gamma_5)
T^\mu_{uu2}(k_1,k_2)(\Cminus-\Cplus\gamma_5),
\eeqa
and
\beqa
M_{11u}^\mu
&=& -N_{11u} (\Cplus-\Cminus\gamma_5)
T^\mu_{11u}(k_1,k_2)(\Cplus+\Cminus\gamma_5), \nn \\
M_{22u}^\mu
&=& -N_{22u} (\Cminus+\Cplus\gamma_5)
T^\mu_{22u}(k_1,k_2)(\Cminus-\Cplus\gamma_5), \nn \\
M_{12u}^\mu
&=& -N_{12u} (\Cplus-\Cminus\gamma_5)
T^\mu_{12u}(k_1,k_2)(\Cminus-\Cplus\gamma_5), \nn \\
M_{21u}^\mu
&=&-N_{21u} (\Cminus+\Cplus\gamma_5)
T^\mu_{21u}(k_1,k_2)(\Cplus+\Cminus\gamma_5).
\eeqa
The crossed amplitudes are related by a change of sign,
and interchange of $k_1$ and $k_2$,
\beqa
\label{EQU:threefour}
M_{uu1}^{\mu\rm (cr)}
&=& N_u\, (\Cplus-\Cminus\gamma_5)
T^\mu_{uu1}(k_2,k_1)(\Cplus+\Cminus\gamma_5), \nn \\
M_{11u}^{\mu\rm (cr)}
&=& N_{11u} (\Cplus-\Cminus\gamma_5)
T^\mu_{11u}(k_2,k_1)(\Cplus+\Cminus\gamma_5),
\eeqa
etc.
If we introduce a sign factor,
\beq
\label{EQU:sign}
S_1=-, \qquad S_2=+,
\eeq
then these results (\ref{EQU:threetwo})--(\ref{EQU:threefour})
can be expressed more compactly as
\beqa
M_{uui}^{\mu}
&=&
-S_{i} N_{u} \left(\Ctilu^{-S_{i}} +S_{i} \Ctilu^{S_{i}}
\gamma_{5}\right)
T^\mu_{uui}(k_1,k_2)
\left( S_{i} \Ctilu^{-S_{i}} - \Ctilu^{S_{i}}\gamma_{5}\right), \nn \\
M_{uui}^{\mu\rm (cr)}
&=&
S_{i} N_{u} \left(\Ctilu^{-S_{i}} +S_{i} \Ctilu^{S_{i}}\gamma_{5}\right)
T^\mu_{uui}(k_2,k_1)
\left(S_{i} \Ctilu^{-S_{i}} - \Ctilu^{S_{i}}
\gamma_{5}\right), \nn \\
M_{iju}^\mu
&=&
-S_{j} N_{iju} \left(\Ctilu^{-S_{i}} +S_{i} \Ctilu^{S_{i}}\gamma_{5}\right)
T^\mu_{iju}(k_1,k_2)
\left(S_{j}\Ctilu^{-S_{j}} - \Ctilu^{S_{j}}
\gamma_{5}\right), \nn \\
M_{iju}^{\mu\rm (cr)}
&=&
S_{j} N_{iju} \left(\Ctilu^{-S_{i}} +S_{i} \Ctilu^{S_{i}}\gamma_{5}\right)
T^\mu_{iju}(k_2,k_1)
\left(S_{j} \Ctilu^{-S_{j}} - \Ctilu^{S_{j}}
\gamma_{5}\right)\;.
\eeqa
Exploiting now the fact that
\beqa
(\Cplus)^2-(\Cminus)^2
&=&2\sin(2\thu), \nn \\
(\Cplus)^2+(\Cminus)^2
&=&2, \nn \\
2\Cplus\Cminus
&=&2\cos(2\thu),
\eeqa
and the expansion (\ref{EQU:triangleadecomp})
in terms of Dirac matrices, we find the structure of $M_{uui}^{\mu}$
to be given by
\beqa
\label{EQU:threeeight}
M_{uui}^{\mu}
&=& 2S_{i}N_{u}E^{a}_{uui\, \alpha}(k_1,k_2) \sin(2\thu)
\gamma^{\alpha}\gamma^{\mu} (g_V^u-g_A^u\gamma_5) \nn \\
& & +2S_{i}N_{u}E^{b}_{uui\, \alpha}(k_1,k_2) \sin(2\thu)
\gamma^{\mu}\gamma^{\alpha} (g_V^u+g_A^u\gamma_5) \nn \\
& & - 2N_{u}O^{a}_{uui}(k_1,k_2) \gamma^{\mu}
\left\{g_V^u +S_{i}\, g_A^u\cos(2\thu)
- \gamma_{5}\left[g_A^u +S_{i}\, g_V^u\cos(2\thu)\right] \right\} \nn \\
& & - 2N_{u}O^{b}_{uui\, \alpha\beta}(k_1,k_2)
\gamma^{\alpha}\gamma^{\mu}\gamma^{\beta}
\left\{g_V^u -S_{i}\, g_A^u\cos(2\thu)
+ \gamma_{5}\left[g_A^u -S_{i}\, g_V^u\cos(2\thu)\right]\right\}\;. \nn \\
\eeqa
Similarly, we find [cf. eq.~(\ref{EQU:trianglebdecomp})]
\beqa
\label{EQU:threenine}
M_{iju}^\mu
&=&
S_{j} N_{iju}
\biggl\{ -S_{j}\Ctilu^{-S_{i}}\Ctilu^{-S_{j}}
+S_{i} \Ctilu^{S_{i}}\Ctilu^{S_{j}} \nn \\
& & \hspace*{30mm} \mbox{}
+ \gamma_{5}\left[ \Ctilu^{-S_{i}}\Ctilu^{S_{j}}
- S_{i}S_{j}\Ctilu^{S_{i}}\Ctilu^{-S_{j}} \right] \biggr\}
 E_{iju}^{\mu}(k_1,k_2) \nn \\
& & -S_{j} N_{iju}\, \gamma^{\alpha}
\biggl\{ S_{j}\Ctilu^{-S_{i}}\Ctilu^{-S_{j}}
+ S_{i}\Ctilu^{S_{i}}\Ctilu^{S_{j}} \nn \\
& & \hspace*{30mm} \mbox{}
- \gamma_{5}\left[ \Ctilu^{-S_{i}}\Ctilu^{S_{j}}
+ S_{i}S_{j} \Ctilu^{S_{i}}\Ctilu^{-S_{j}}\right] \biggr\}
O_{iju\, \alpha}^{\mu}(k_1,k_2)
\eeqa

{}From eq.~(\ref{EQU:eqtwosix}), we define $M^\mu$ by
\beqa
\label{EQU:gmu}
\tilde G^\mu
&\equiv&\sum_{\mbox{\scriptsize generations}}
\bigl(\tilde G^\mu_u + \tilde G^\mu_d \bigr) \cr
&=&\bar u(k_2) \Bigl[M^\mu(k_1,k_2)
-C^{-1}M^{\mu{\rm (cr)T}}(k_1,k_2)C\Bigr]
C^{-1}\bar u^{\rm T}(k_1).
\eeqa
Thus, when summed over flavours [cf.\ eq.~(\ref{EQU:glueu})], we have
\beqa
M^\mu
&=&\sum_{\mbox{\scriptsize generations}} \Bigl\{
  \left(M^\mu_{uu1} + M^\mu_{uu2}\right)
+ \left(M_{11u}^{\mu} + M_{22u}^{\mu}
+       M_{12u}^{\mu} + M_{21u}^{\mu}\right) \nn \\
& &\phantom{\sum_{\mbox{\scriptsize generation}}}
+ \left(M^\mu_{dd1} + M^\mu_{dd2}\right)
+ \left(M_{11d}^{\mu} + M_{22d}^{\mu}
+       M_{12d}^{\mu} + M_{21d}^{\mu}\right) \Bigr\}
\eeqa
and a similar expression $M^{\mu{\rm (cr)}}$ for the crossed amplitudes.

Using eqs.~(\ref{EQU:threeeight}) and~(\ref{EQU:threenine}),
we get the following structure in terms of Dirac matrices
\beqa
\label{EQU:MMM}
M^{\mu} - C^{-1}M^{\mu{\rm (cr)T}}C
&=&
\left( {\cal V}_{\alpha}^{a} + \gamma_{5}{\cal A}_{\alpha}^{a} \right)
\gamma^{\alpha}\gamma^{\mu}
+ \left( {\cal V}_{\alpha}^{b} + \gamma_{5}{\cal A}_{\alpha}^{b} \right)
\gamma^{\mu}\gamma^{\alpha} \nn \\
& & + \left( {\cal V}^{c} + \gamma_{5}{\cal A}^{c} \right) \gamma^{\mu}
+ \left( {\cal V}_{\alpha\beta}^{d}
+ \gamma_{5}{\cal A}_{\alpha\beta}^{d} \right)
\gamma^{\alpha}\gamma^{\mu}\gamma^{\beta} \nn \\
& & + {\cal V}^{e\mu} + \gamma_{5}{\cal A}^{e\mu}
+ \left( {\cal V}_{\alpha}^{f\mu}
+ \gamma_{5}{\cal A}_{\alpha}^{f\mu} \right) \gamma^{\alpha} \; .
\eeqa
The ${\cal V}^{c}$ and ${\cal V}_{\alpha}^{f\mu}$ contributions vanish since
two Majorana fermions cannot form a vector current:
\beq
\renewcommand{\arraystretch}{1.0}
\overline{\Psi}_{g}\gamma^{\mu}\Psi_{g}
= \left(\gli \quad \ogli\right)
 \left( \begin{array}{cc} 0 & \sigma^{\mu} \\
\overline{\sigma}^{\mu} & 0 \end{array} \right)
 \left( \begin{array}{c} \gli \\ \ogli \end{array} \right)
= \ogli\overline{\sigma}^{\mu}\gli + \gli\sigma^{\mu}\ogli = 0.
\renewcommand{\arraystretch}{2.0}
\eeq
The other ${\cal V}$ and ${\cal A}$ terms are given in Appendix~A.
All the remaining ${\cal V}$
and also the pseudoscalar ${{\cal A}}^{e\mu}$ vanish,
and eq.~(\ref{EQU:MMM}) takes the simple form
\beqa
\label{EQU:MMMsimple}
M^{\mu} - C^{-1}M^{\mu{\rm (cr)T}}C
&=&
  {\cal A}_{\alpha}^{a} \gamma^{\alpha}\gamma^{\mu}\gamma_{5}
+ {\cal A}_{\alpha}^{b} \gamma^{\mu}\gamma^{\alpha}\gamma_{5}
- {\cal A}^{c} \gamma^{\mu}\gamma_{5} \nn \\
& &
- {\cal A}_{\alpha}^{f\mu} \gamma^{\alpha}\gamma_{5}
- {\cal A}_{\alpha\beta}^{d}
\gamma^{\alpha}\gamma^{\mu}\gamma^{\beta}\gamma_{5} \; .
\eeqa
%

%%%%%%%%%%%%%%%%%%%%%%%%%%%%%%%%%%%%%%%%%%%%%%%%%%%%%%%%%%%%%%%%%%%%%%%%%%%
\section{The Cross Section}
\setcounter{equation}{0}
\label{:spinsum}
Evaluating the spin sum, we get [cf.\ eq.~(\ref{EQU:amplitude})]
\beqa
\label{EQU:spinsum}
X
&=& {1\over4} \sum_{\mbox{{\scriptsize spin}}}
{\cal M}^{\dagger}{\cal M} \nn \\
&=& {1\over4} \sum_{\mbox{{\scriptsize spin}}}
\left({L}^{\mu}D_{F\,\mu\nu}{\Glue}^{\nu}\right)
\left(\Glue^{\alpha\,\dagger}D^\dagger_{F\,\alpha\beta}
L^{\beta\,\dagger}\right)
\nn \\
&=& \frac{g^{2}(g_V^2+g_A^2)}{4\cos^{2}\thW} \;
\frac{1}
{\left(s^2 - \mZ^{2} \right)^{2}}\left\{ p_{1\mu}p_{2\nu}
+ p_{1\nu}p_{2\mu}
- (p_{1}\cdot p_{2})g_{\mu\nu} \right\} {\cal T}^{\mu\nu},
\eeqa
where [cf.\ eq.~(\ref{EQU:gmu})]
\beqa
\label{EQU:Ttilde}
{\cal T}^{\mu\nu}
&=&
\sum_{\mbox{{\scriptsize spin}}}
\bar u(k_2)
\left[M^{\mu}(k_1,k_2) - C^{-1} M^{\mu\,{\rm (cr) T}}(k_1,k_2)C\right]
C^{-1} \bar u^{\rm T}(k_1) \nn \\
& & \times
u^{\rm T}(k_1)\gamma^{0\Tsp}(-C^{-1})\left[M^{\nu\,\dagger}(k_1,k_2)
- C M^{\nu\,{\rm (cr) T\dagger}}(k_1,k_2)C^{-1} \right]
\gamma^{0}u(k_2) \nn \\
&=& \Tr\biggl[
\left\{M^{\mu}(k_1,k_2)
- C^{-1} M^{\mu\,{\rm (cr) T}}(k_1,k_2)C\right\}
(\not\!k_{1}-\mgli) \nn \\
& & \hspace*{8mm} \times\gamma^{0}\left\{M^{\nu\,\dagger}(k_1,k_2)
 - C M^{\nu\,{\rm (cr) T\dagger}}(k_1,k_2)C^{-1} \right\}
\gamma^{0}(\not\!k_{2}+\mgli)\biggr].
\eeqa
We have here used $\gamma^{\mu\Tsp} = - C\gamma^{\mu}C^{-1}$.

Invoking eq.~(\ref{EQU:MMMsimple}), we obtain the structure
of the tensor ${\cal T}^{\mu\nu}$ in terms of Dirac matrices as
\beqa
\label{EQU:CalT}
{\cal T}^{\mu\nu}
&=&-\Tr\Biggl[
\biggl\{
  {\cal A}_{\alpha}^{a} \gamma^{\alpha}\gamma^{\mu}\gamma_{5}
+ {\cal A}_{\alpha}^{b} \gamma^{\mu}\gamma^{\alpha}\gamma_{5}
- {\cal A}^{c} \gamma^{\mu}\gamma_{5}
- {\cal A}_{\alpha}^{f\mu} \gamma^{\alpha}\gamma_{5} \nn \\
& & \hspace*{2mm}\mbox{}
- {\cal A}_{\alpha\beta}^{d}
\gamma^{\alpha}\gamma^{\mu}\gamma^{\beta}\gamma_{5}
\biggr\}
(\not\!k_{1}-\mgli) \nn \\
& & \hspace*{2mm}\mbox{}\times\biggl\{
 {\cal A}_{\rho}^{a\dagger} \gamma^{\nu}\gamma^{\rho}\gamma_{5}
+ {\cal A}_{\rho}^{b\dagger} \gamma^{\rho}\gamma^{\nu}\gamma_{5}
+ {\cal A}^{c\dagger} \gamma^{\nu}\gamma_{5}
+ {\cal A}_{\rho}^{f\nu\dagger} \gamma^{\rho}\gamma_{5} \nn \\
& & \hspace*{2mm}\mbox{}
+ {\cal A}_{\rho\sigma}^{d\dagger}
\gamma^{\sigma}\gamma^{\nu}\gamma^{\rho}\gamma_{5}
\biggr\}
(\not\!k_{2}+\mgli)
\Biggr], \nn \\
\eeqa
and evaluate the trace using computer algebra \cite{REDUCE,MAPLE}.

By summing over the eight gluino colours,
and integrating over the solid angle, we find that the cross section
is proportional to the square of the sum of
two partial amplitudes,
corresponding to the contributions of the two diagrams (a) and (b).
This is possible, since by general arguments \cite{Nelson},
there is essentially only one invariant amplitude.
The integrated cross section thus takes the form
\beq
\label{EQU:sigma}
\sigma
= \frac{g^{2}\pi^{3}\left(g_{V}^{2}+g_{A}^{2}\right)
\left(\sqrt{E^2-\mgli^{2}\,}\,\right)^{3}}{12E\cos^{2}\thW
\left[ \left(s - \mZ^{2} \right)^2
+ \Gamma_Z^{2}\mZ^{2} \right]} \;
\left|\sum({\cal A}_a+{\cal A}_b)\right|^2,
\eeq
with $E$ the beam energy and
the sum running over quark flavours $q$.
The two partial amplitudes correspond to diagrams (a) and (b) and are
given as
\beqa
\label{EQU:eqampl}
{\cal A}_a
&=& 4\sum_{i}S_{i}N_q
\biggl\{ F^{00}_{qqi}\left( 2\hat b_q\mgli m_q
+ f_{qqi}\mgli^{2} - v_{qqi} m_q^{2} \right)
- 4F^{01}_{qqi}\mgli\left(\hat b_q m_q+f_{qqi}\mgli \right) \nn \\
& & \hspace*{6mm}
+ 2F^{02}_{qqi}f_{qqi}\mgli^{2}
- 2F^{11}_{qqi}f_{qqi}\left({1\over2}\,s -\mgli^{2}\right)
+ G_{qqi}f_{qqi}  \biggr\},
\nn \\
{\cal A}_b
&=& 4\sum_{i,j}S_{j}N_{ijq}b_{ijq}G_{ijq},
\eeqa
with $S_i$ the sign factor of eq.~(\ref{EQU:sign})
and the dependence on the electroweak and chiral mixing angles
given by the coefficients
\beqa
\hat{b}_{q}
&=& -2g_A^{q}\sin(2\tq), \nn \\
b_{ijq}
&=& \Ctilq^{-S_{i}}\Ctilq^{S_{j}}
+ S_{i}S_{j} \Ctilq^{S_{i}}\Ctilq^{-S_{j}}, \nn \\
f_{qqi}
&=& 2S_{i}\left\{g_A^q - S_{i}\, g_V^q\cos(2\tq)\right\}, \nn \\
v_{qqi}
&=& -2S_{i}\left\{g_A^q + S_{i}\, g_V^q\cos(2\tq)\right\},
\eeqa
which are read off from the contributing ${\cal A}$ terms
of eq.~(\ref{EQU:CalT}).
We note that the amplitudes (\ref{EQU:eqampl}) contain terms
that apparently are odd in the masses, i.e., proportional to
$\mgli m_q$.
These arise from the chiral mixing, i.e., they are multiplied by
factors $\hat b_q$ which also are odd in these masses, and
vanish in the limit of no mixing.
The integrals $F_{qqi}^{ab}$, $G_{qqi}$ and $G_{ijq}$ are given
in Appendix~A.

The above result, eq.~(\ref{EQU:sigma}), is given as an integrated
cross section.  Actually, since there is only one invariant
amplitude, whose structure is determined by the fact that it
describes the annihilation of two massless fermions to a pair
of self-conjugate fermions \cite{Nelson}, the angular distribution
is given by the familiar expression
\beq
{\dd\sigma\over\dd\Omega}
={3\over 16\pi}\,\sigma\,(1+\cos^2\theta).
\eeq
%

%%%%%%%%%%%%%%%%%%%%%%%%%%%%%%%%%%%%%%%%%%%%%%%%%%%%%%%%%%%%%%%%%%%%%%%%%%%
\section{Results}
\setcounter{equation}{0}
\label{:results}

In order to better understand what is required for the cross
section to be large, let us first state the conditions that must
be satisfied in order for it to vanish.

\medskip\noindent{\bf Conditions for vanishing cross section}\\
The gluino pair production cross section would {\it vanish}
if the following
{\it two conditions were both satisfied}. These conditions are
\cite {Campbell2}

\begin{enumerate}
\item mass degeneracy in each quark isospin doublet,
$m_d=m_u$ (this is violated),

\item mass degeneracy in each squark isospin doublet, i.e.,
$m_{{\tilde d}_1}=m_{{\tilde d}_2}
=m_{{\tilde u}_1}=m_{{\tilde u}_2}$, for each generation.

\end{enumerate}
Kane and Rolnick \cite{Kane} state that in the case of $Z$ decay,
the cross section vanishes when $m_{{\tilde q}}=m_q$.
We do not reproduce this requirement, but instead the conditions
(1) and (2) above.

For comparison, in the case of no axial coupling to the $Z$,
i.e., in the QED limit,
the cross section would {\it vanish} if there is \cite{Nelson}

\begin{itemize}
\item mass degeneracy in each squark chiral doublet,
i.e., $m_{\tilde u1}=m_{\tilde u2}$, and $m_{\tilde d1}=m_{\tilde d2}$
for each generation.
This condition is less strong than item~(2) above.
\end{itemize}

The magnitude of the cross section will depend on how strongly
these conditions (1) and (2) are violated.
Especially for the third generation, item~(1) is violated.
This is generally believed to imply that the squark isospin
doublets are not degenerated either.
However, in a consistent MSSM, the squark masses
can not be specified as free parameters, they emerge as dependent
on the more fundamental parameters of the Lagrangian.
Furthermore, there are four squark masses for each generation.
It is therefore not possible to make simple (and correct) statements
about the magnitude of the cross section.

For the purpose of developing some intuition for how large
the gluino pair production cross section would be at LEP,
we show in figure~3 the ratio
\beq
R=\frac{\sigma(e^+e^-\to\gluino\gluino)}{\sigma(e^+e^-\to\mu^+\mu^-)}
\eeq
vs.\ maximal squark mass splitting $\delta \msquark$.
The plot is based on a scan of the MSSM parameter space,
at grid points given by
\beqa
\label{EQU:paramspace}
\tan\beta &\in& \{1.1,5,15,30\}, \nn \\
\mu &\in& \{0,\pm20,\pm40,\pm70,\pm100,\pm200,\pm300,\pm500\}~\GeV, \nn \\
A_t &\in& \{0,10,20,40,70,100,200,300,500,800,1000\}~\GeV, \nn \\
A_b &\in& \{0,10,20,40,70,100,200,300,500,800,1000\}~\GeV, \nn \\
\MsQT      &\in& \{0,10,20,40,70,100,200,300,500,800,1000\}~\GeV, \nn \\
{\MsT} &\in& \{0,10,20,40,70,100,200,300,500,800,1000\}~\GeV, \nn \\
{\MsB} &\in& \{0,10,20,40,70,100,200,300,500,800,1000\}~\GeV,
\eeqa
for gluino, bottom and top quark masses given by the
``standard values",
\beq
\label{EQU:standard}
\mgluino=3.5~\GeV, \qquad m_b=4.8~\GeV, \qquad m_t=170~\GeV.
\eeq
We here consider only the contributions from the third generation,
so $t$ (or $T$) and $b$ (or $B$) refer to $u$ (or $U$) and $d$ (or $D$)
in the Lagrangian (\ref{EQU:Lagrangeone})--(\ref{EQU:Lagrangefour}).
All encountered cross section ratios lie in the light shaded region, where
the horizontal axis gives the largest
resulting squark mass difference,
$\delta \msquark=\max_{i,j}|m_{\tilde q_i}-m_{\tilde q'_j}|$.
No values are found within the dark shaded or the white regions.
The cross section ratios are thus typically
between $10^{-5}$ and $10^{-2}$.
(The $Z$ branching ratio is obtained upon multiplying by 3.3\%.)
The jagged borders are ascribed
to the discreteness of the sampling, as well as the rather complex dependence
the cross section has on the many parameters.  Parameter sets that lead to
any one of the squarks being light, $\msquark<45~\GeV$, are left out,
since such light squarks would have been detected at LEP
\cite{LEPsquark}.

The value for the gluino mass, $\mgluino=3.5~\GeV$, has been
chosen as representative of the ``light-gluino window".
Actually, the cross section has only a very weak dependence
on the gluino mass, as long as it is well below
the kinematical threshold \cite{Kilengthesis}.

\medskip\noindent{\bf Dependence on squark and top masses}\\
As noted previously \cite{Kane},
the gluino cross section tends to increase with increasing top mass,
but the way it increases depends on the other parameters.
This is illustrated in figure~4, where we show the ratio $R$ as a function
of stop mass (denoted $\msu$), for different values of top mass
(denoted $m_u$).
However, this figure is somewhat idealized in the sense that
the squark masses are set by hand, they do not result naturally
from some set of fundamental parameters of the Lagrangian.
Two sets of parameters are considered, each set is for
$m_{\tilde t_1}=m_{\tilde t_2}\equiv m_{\tilde u}$ and
$m_{\tilde b_1}=m_{\tilde b_2}\equiv m_{\tilde d}$.
The three steep curves are for $m_{\tilde d}=50~\GeV$,
whereas the other three are for $m_{\tilde d}=200~\GeV$.
For each set, three values of the top quark
mass are considered, $m_t=0$, 50, and 170~GeV.
We note that if $m_u=m_d$ ($=0$ GeV) and
$m_{\tilde u}=m_{\tilde d}$, then the cross section vanishes,
in accordance with items (1) and (2) above.

\medskip\noindent{\bf Dependence on $\tan\beta$ and $\mu$}\\
We can now start to address the question of which parameters would
be restricted by an experimental limit on the cross section.
With the parameters $A_t$, $A_b$, $\MsQT$, $\MsT$, and $\MsB$
allowed to take on values in the set
\beqa
A_t &\in& \{0,10,20,50,100,300,500,800\}~\GeV, \nn \\
A_b &\in& \{0,10,20,50,100,300,500,800\}~\GeV, \nn \\
\MsQT &\in& \{0,10,20,50,100,300,500,800\}~\GeV, \nn \\
{\MsT} &\in& \{0,10,20,50,100,300,500,800\}~\GeV, \nn \\
{\MsB} &\in& \{0,10,20,50,100,300,500,800\}~\GeV,
\eeqa
we have scanned for extrema of the gluino cross section as a
function of $\tan\beta$ and $\mu$.
It turns out that there is little dependence on the latter parameters.
In fact, the minimal values found are $R_{\min{}} \simeq 10^{-6}$, whereas
the maximal values are $R_{\max{}}\simeq0.01$--0.02, with a rather weak
dependence on $\tan\beta$ and $\mu$, for
$1.1\le\tan\beta\le50$ and $|\mu|\le500~\GeV$.
{\it Thus, an upper limit on the gluino pair production cross section
does not significantly restrict neither $\tan\beta$ nor $\mu$.}

The lightest squark, which is the lightest stop, $\tilde t_2$,
will exceed about 350~GeV
for the values of $A_t$, $A_b$, $\MsQT$, ${\MsT}$ and ${\MsB}$
which minimize $R$,
in the given range of $\tan\beta$ and $\mu$.

\medskip\noindent{\bf Dependence on $\tan\beta$ and $\MsQT$}\\
In figure~5 we indicate the dependence of the cross section
on $\tan\beta$ and $\MsQT$, for the following choice of the other
parameters,
\beqa
\label{EQU:masses}
\MsQT &=&{\MsT}={\MsB}, \nn \\
A     &=&A_t=A_b,
\eeqa
with
\beqa
\mu &\in& \{0,\pm20,\pm40,\pm70,\pm100,\pm200,\pm300,\pm500\}~\GeV, \nn \\
A   &\in& \{0,10,20,40,70,100,200,300,500,800,1000\}~\GeV
\eeqa
and for the ``standard values" for gluino, $b$ and $t$ quark masses
given by eq.~(\ref{EQU:standard}).
Clearly, an upper bound on the cross section ratio of e.g. $10^{-3}$,
would rule out values of $\MsQT$ below about 350~GeV.

A lower bound on $\MsQT$ would also lead to a lower bound
on the heaviest squark (for this set of parameters,
always the heaviest stop, $\tilde t_1$)
about similar in magnitude to $\MsQT$ \cite{Kilengthesis}.

\medskip\noindent{\bf Dependence on $\MsB$ and $\MsT$}\\
The correlation between the cross section ratio $R$ and $\MsQT$
is however not quite as simple as that shown in figure~5 if we
relax the condition (\ref{EQU:masses}).
It turns out that $R$ can become larger than $10^{-3}$ even for rather
low values $\MsT\leq 100~\GeV$, {\it provided $\MsB$ is high}.
This is illustrated in figure~6, where we show regions
in the $\MsB$--$\MsT$ plane where $R$ exceeds $10^{-3}$ for
given upper bounds on $\MsQT$. Two cases are considered,
$R_{\min{}}$ in (a), and $R_{\max{}}$ in (b),
where ``min" and ``max" refer to scans over the parameter values
\beqa
\tan\beta &\in&  \{ 1.1,5,15,30\}, \nn \\
\mu       &\in&  \{0,\pm50,\pm100,\pm200,\pm500\}~\mbox{GeV}, \nn \\
A_{t}     &\in&  \{0,10,20,50,100,300,800\}~\mbox{GeV}, \nn \\
A_{b}     &\in&  \{0,10,20,50,100,300,800\}~\mbox{GeV}.
\eeqa
The gluino and quark masses considered are the ``standard values",
and for $\MsQT$ we have taken
\beq
\MsQT  \in  \{50,100,300,800\}~\mbox{GeV}.
\eeq
%
%%%%%%%%%%%%%%%%%%%%%%%%%%%%%%%%%%%%%%%%%%%%%%%%%%%%%%%%%%%%%%%%%%%%%%%%%%%
\section{Discussion}
\setcounter{equation}{0}
\label{:discussion}
The present study does not address the question of decay or fragmentation.
In order to consider a ``worst case" scenario,
we basically assume the gluinos are stable and form gluinoballs.
If they are unstable and decay, detection would be easier.
These gluinoballs must be colour singlets,
but could be electrically charged, in which
case they would show up in the detectors, or neutral, in which case they
would presumably escape undetected.
{\it However, in the latter case, since
they are produced far above threshold, one would expect a few ordinary
hadrons (e.g., pions) to also emerge from the fragmentation process.}
These would be detected, and give standard SUSY-triggers of
considerable missing energy.

For the sake of definiteness, suppose one can rule out the production
of gluino pairs at a level of at most 10 events
per 1 million $Z$ decays.
This would imply $R<10^{-5}/3.3\%$, or $R<3\cdot10^{-4}$.
It follows from figure~3 that this condition would exclude much
of the ``Physical Region".
{}From figures~5 and 6 we see that lower limits on the
soft-supersymmetry-breaking parameters would be obtained, but that
the precise limits would depend on whether these parameters are
related.

In summary, the pair production of gluinos, without accompanying
quark jets, is in $Z$ decay large enough to be measurable in
much of the MSSM parameter space, and should therefore
be searched for vigorously.

%%%%%%%%%%%%%%%%%%%%%%%%%%%%%%%%%%%%%%%%%%%%%%%%%%%%%%%%%%%%%%%%%%%%%%%%
\medskip
It is a pleasure to thank T. Medcalf and F. Richard for useful comments.
This research has been supported by the Research Council of Norway.
%%%%%%%%%%%%%%%%%%%%%%%%%%%%%%%%%%%%%%%%%%%%%%%%%%%%%%%%%%%%%%%%%%%%%%%%
\newpage
\appendix
\section*{Appendix A}
\setcounter{equation}{0}
% CERNTEX does not like the next two commands:
%\def\thesection{{\rm A}}
%\renewcommand{\theequation}{\thesection.\arabic{equation}}
\renewcommand{\thesection}{A}
This appendix provides some information on the triangle integrals.

\smallskip
\noindent
{\bf The integrals of eqs.~(\ref{EQU:triangleadecomp})
and (\ref{EQU:trianglebdecomp}):} \\
The quantities appearing in eqs.~(\ref{EQU:triangleadecomp})
and (\ref{EQU:trianglebdecomp})
can be expressed in terms of more basic integrals as
\beqa
\label{EQU:uyhgh}
E^{a\mu}_{uui}(k_1,k_2)
&=&
-i\pi^{2}\left[
k_{1}^{\mu}F^{01}_{uui}
+k_{2}^{\mu}\left( F^{00}_{uui} - F^{10}_{uui} \right)
\right], \nn \\
E^{b\mu}_{uui}(k_1,k_2)
&=&
i\pi^{2}\left[
k_{1}^{\mu} \left( F^{00}_{uui} - F^{01}_{uui} \right)
+ k_{2}^{\mu} F^{10}_{uui}
\right], \nn \\
O^{b\mu\nu}_{uui}(k_1,k_2)
&=&
i\pi^{2} \biggl[
{1\over 4}g^{\mu\nu} \left( {1\over\eps}-\gamma+2 - 2G_{uui} \right)
+ k_{1}^{\mu}k_{1}^{\nu} \left( F^{01}_{uui} - F^{02}_{uui} \right)
\nn \\
& & \hspace*{7mm}\mbox{}
+ k_{1}^{\mu}k_{2}^{\nu} F^{11}_{uui}
+ k_{2}^{\mu}k_{1}^{\nu}
\left( F^{00}_{uui} - F^{10}_{uui} + F^{11}_{uui} - F^{01}_{uui} \right)
+ k_{2}^{\mu}k_{2}^{\nu} \left( F^{10}_{uui} - F^{20}_{uui} \right)
\biggr], \nn \\
O^{a}_{uui}(k_1,k_2)
&=& - i\pi^{2}F^{00}_{uui}\; , \nn \\
E_{iju}^{\mu}(k_1,k_2)
&=& i\pi^{2}
\left[ k_{1}^{\mu} \left( 2F^{01}_{iju} - F^{00}_{iju} \right)
      -k_{2}^{\mu} \left( 2F^{10}_{iju} - F^{00}_{iju} \right)
\right], \nn \\
O_{iju}^{\mu\nu}(k_1,k_2)
&=&
i\pi^{2}
\biggl[
{1\over 2}g^{\mu\nu}\left( {1\over\eps} - \gamma +2 - 2G_{iju} \right)
+ k_{1}^{\mu}k_{1}^{\nu} \left( F^{01}_{iju} - 2F^{02}_{iju} \right)
+ k_{1}^{\mu}k_{2}^{\nu} \left( 2F^{11}_{iju} - F^{01}_{iju} \right)
\nn \\
& & \hspace*{7mm}\mbox{}
+ k_{2}^{\mu}k_{1}^{\nu} \left( 2F^{11}_{iju} - F^{10}_{iju} \right)
+ k_{2}^{\mu}k_{2}^{\nu} \left( F^{10}_{iju} - 2F^{20}_{iju}\right)
\biggr],
\eeqa
with $1/\epsilon$ representing the UV-divergent part,
$\gamma$ the Euler constant, and
the integrals over Feynman parameters defined by
\beqa
\label{EQU:eqthreesix}
F^{ab}_{qqi}
&=& \int_{0}^{1}\dd x\;
\int_{0}^{1-x}\dd z\;\frac{z^{a}x^{b}}{h_{qqi}}, \nn \\
F^{ab}_{ijq}
&=& \int_{0}^{1}\dd x\;
\int_{0}^{1-x}\dd z\;\frac{z^{a}x^{b}}{h_{ijq}}, \nn \\
G_{qqi}
&=& \int_{0}^{1}\dd x\;\int_{0}^{1-x}\dd z\;
\log \frac{h_{qqi}}{\mu^{2}} , \nn \\
G_{ijq}
&=& \int_{0}^{1}\dd x\;\int_{0}^{1-x}\dd z\;
\log \frac{h_{ijq}}{\mu^{2}} ,
\eeqa
with
\beqa
h_{qqi}
&=& \mgli^{2}(x+z)(x+z-1) - sxz
- (\Msqa{i}^{2}-m_q^{2})(x+z) + \Msqa{i}^{2} -\ieps, \nn \\
h_{ijq}
&=& \mgli^{2}(x+z)(x+z-1) - sxz + (\Msqa{j}^{2} - m_q^{2})x \nn \\
& &
+ (\Msqa{i}^{2}-m_q^{2})z + m_q^{2} -\ieps.
\eeqa

The parameter $\mu$ is a renormalization mass.
When the amplitude is summed over both members of an
isospin doublet, the $\mu$-dependence cancels.
The integrals $F^{ab}_{qqi}$, $G_{qqi}$ and $G_{ijq}$ can be
evaluated in terms of dilogarithms
($F^{ab}_{ijq}$ does not contribute).
Performing the integration over $z$, we find that
the $F^{ab}_{qqi}$, $G_{qqi}$ and $G_{ijq}$ can be expressed
in terms of the one-dimensional integrals
\beqa
I^{m}
&=& \int_{0}^{1}\mbox{\rm d}x\,
\frac{x^{m}\log\left[ax + b \pm \sqrt{c(x^{2} +2dx + e)\;}
\pm i\eps \right]}{\sqrt{c(x^{2} +2dx + e)\;}} \nn \\
J
&=& \int_{0}^{1}\mbox{\rm d}x\;
\sqrt{c\left(x^{2} + 2dx + e \right)}\;\log\left[ax + b
\pm \sqrt{c\left(x^{2} + 2dx + e \right)}\; \pm i\eps \right] \nn \\
K^{n}
&=& \int_{0}^{1}\mbox{\rm d}x\;
x^{n}\log\left[(ax + b)^{2} - c(x^{2} +2dx + e) - i\eps \right] .
\eeqa
Here $m=0,1,2$, and $n=0,1$.
The $K^{n}$ integral is straightforward.
The arguments of the square roots in $I^{m}$ and $J$ may
change sign within the domain of integration.
The $I^{m}$ and $J$ integrals are evaluated using
the following substitutions
\beq
\renewcommand{\arraystretch}{1.0}
x=y-d \rightarrow y=u\sqrt{d^{2}-e\,}\; \rightarrow
\left\{ \begin{array}{ll}
u=\cosh\alpha \rightarrow v=\tanh(\alpha/2) & \mbox{when $u\geq1$}, \\
u=-\cosh\alpha \rightarrow v=\tanh(\alpha/2) & \mbox{when $u\leq1$}, \\
u=\sin\alpha \rightarrow v=\tan(\alpha/2) & \mbox{when $|u|<1$} .
\end{array} \right.
\renewcommand{\arraystretch}{2.0}
\eeq

The integrals $F^{ab}_{uui}$ and $F^{ab}_{iju}$ satisfy the symmetry
relations
\beq
F^{ab}_{uui}
= F^{ba}_{uui} \quad \mbox{and} \quad F^{ab}_{ijq}
= F^{ba}_{jiq},
\label{fghdfdf}
\eeq
This is easily checked by interchanging the parametric integrations.

%The only remaining contribution from $T^\mu_{iju}(k_1,k_2)$
%to the tensor ${\cal T}^{\mu\nu}$ will be the $G_{iju}$ coming
%from $O_{iju}^{\mu\nu}$.

\medskip
\noindent
{\bf The integrals of eq.~(\ref{EQU:MMM}):} \\
The ${\cal V}$ integrals are defined by
\beqa
{\cal V}_{\alpha}^{a}
&=& 2\sum_{iq}S_{i}\, N_{q}\, g_V^q\sin(2\tq)
\left[ E^{a}_{qqi\,\alpha}(k_1,k_2)
+ E^{b}_{qqi\,\alpha}(k_2,k_1) \right], \nn \\
{\cal V}_{\alpha}^{b}
&=& 2\sum_{iq}S_{i}\, N_{q}\, g_V^q\sin(2\tq)
\left[ E^{a}_{qqi\,\alpha}(k_2,k_1)
+ E^{b}_{qqi\,\alpha}(k_1,k_2) \right], \nn \\
{\cal V}_{\alpha\beta}^{d}
&=& - 2\sum_{iq}N_{q}
\left[O^{b}_{qqi\, \alpha\beta}(k_1,k_2)
- O^{b}_{qqi\, \beta\alpha}(k_2,k_1)\right]
\left[g_V^q -S_{i}\, g_A^q\cos(2\tq)\right], \nn \\
{\cal V}^{e\mu}
&=& -\sum_{ijq}S_{j} N_{ijq}
\left[ E_{ijq}^{\mu}(k_1,k_2) + E_{ijq}^{\mu}(k_2,k_1) \right]
\left( S_{j}C_{ijq}^{-S_{i}}C_{ijq}^{-S_{j}}
-S_{i} C_{ijq}^{S_{i}}C_{ijq}^{S_{j}}\right).
\eeqa
Using eqs.~(\ref{EQU:uyhgh}) and~(\ref{fghdfdf}),
all these terms can be shown to vanish,
i.e., we are left with only the ${\cal A}$ type terms,
\beqa
{\cal A}_{\alpha}^{a}
&=& -2\sum_{iq}S_{i}\, N_{q}\, g_A^q\sin(2\tq)
\left[ E^{a}_{qqi\,\alpha}(k_1,k_2)
- E^{b}_{qqi\,\alpha}(k_2,k_1) \right], \nn \\
{\cal A}_{\alpha}^{b}
&=& -2\sum_{iq}S_{i}\, N_{q}\, g_A^q\sin(2\tq)
\left[ E^{a}_{qqi\,\alpha}(k_2,k_1)
- E^{b}_{qqi\,\alpha}(k_1,k_2) \right], \nn \\
{\cal A}^{c}
&=& - 2\sum_{iq}N_{q}
\left[O^{a}_{qqi}(k_1,k_2)+O^{a}_{qqi}(k_2,k_1)\right]
\left[g_A^q +S_{i}\,g_V^q\cos(2\tq)\right], \nn \\
{\cal A}_{\alpha\beta}^{d}
&=& 2\sum_{iq}N_{q}
\left[O^{b}_{qqi\, \alpha\beta}(k_1,k_2)
+O^{b}_{qqi\, \beta\alpha}(k_2,k_1)\right]
\left[g_A^q -S_{i}\, g_V^q\cos(2\tq)\right], \nn \\
{\cal A}^{e\mu}
&=&  \sum_{ijq}S_{j} N_{ijq}
\left[ E_{ijq}^{\mu}(k_1,k_2) + E_{ijq}^{\mu}(k_2,k_1) \right]
\left( C_{ijq}^{-S_{i}}C_{ijq}^{S_{j}}
- S_{i}S_{j}C_{ijq}^{S_{i}}C_{ijq}^{-S_{j}} \right), \nn \\
{\cal A}_{\alpha}^{f\mu}
&=& - \sum_{ijq}S_{j} N_{ijq}
\left[ O_{ijq\, \alpha}^{\mu}(k_1,k_2)
+ O_{ijq\, \alpha}^{\mu}(k_2,k_1) \right]
\left( \Ctilq^{-S_{i}}\Ctilq^{S_{j}}
+ S_{i}S_{j} \Ctilq^{S_{i}}\Ctilq^{-S_{j}}\right). \nn \\
\eeqa
Using eqs.~(\ref{EQU:uyhgh}) and~(\ref{fghdfdf}),
we find that ${{\cal A}}^{e\mu}$ vanishes, and eq.~(\ref{EQU:MMM})
reduces to eq.~(\ref{EQU:MMMsimple}).
%%%%%%%%%%%%%%%%%%%%%%%%%%%%%%%%%%%%%%%%%%%%%%%%%%%%%%%%%%%%%%%%%%%%%%%%%%%

\newpage
%%%%%%%%%%%%%%%%%%%%%%%%%%%%%%%%%%%%%%%%%%%%%%%%%%%%%%%%%%%%%%%%%%%%%%%%

%%%%%%%%%%%%%%%%%%%%%%%%%%%%%%%%%%%%%%%%%%%%%%%%%%%%%%%%%%%%%%%%%%%%%%%%
\renewcommand{\arraystretch}{1.5}
\clearpage
\centerline{\bf Figure captions}

\vskip 15pt
\def\fig#1#2{\hangindent=.65truein \noindent \hbox to .65truein{Fig.\ #1.
\hfil}#2\vskip 2pt}

\fig1{The two classes of Feynman diagrams for $e^{+}\,e^{-} \rightarrow
\tilde{g}\,\tilde{g}$.}

\fig2{The couplings involved in the process
      $e^{+}\,e^{-} \rightarrow \tilde{g}\,\tilde{g}$.}

\fig3{Cross section ratios $R=\sigma(e^+e^-\to\gluino\gluino)/
\sigma(e^+e^-\to\mu^+\mu^-)$ at the $Z$ resonance.
The figure shows the result of a scan of parameter space,
eq.~(\ref{EQU:paramspace}), against the {\it largest} resulting
squark mass difference.}

\fig4{Cross section ratios $R$ vs. stop mass
$m_{{\tilde t}_1}=m_{{\tilde t}_2}\equiv m_{{\tilde u}}$.
Two values of sbottom mass are considered,
$m_{{\tilde b}_1}=m_{{\tilde b}_2}\equiv m_{{\tilde d}}=50$~GeV
and 200~GeV, together with three values of ($u$ or) top quark mass.}

\fig5{Regions of {\it lower} bounds on $R$ in the plane spanned
by the soft squark mass parameter $\MsQT$
[cf.\ eqs.~(\ref{EQU:Lagrangefour}) and (\ref{EQU:massnomix})]
and $\tan\beta$.
A somewhat special case is considered, cf.\ eq.~(\ref{EQU:masses}).
We here consider the values of $\mgluino$, $m_b$, and $m_t$
given by eq.~(\ref{EQU:standard}).}

\fig6{Regions where $R\geq 10^{-3}$ are
for different values of $\MsQT$ outlined
in the plane spanned by $\MsT$ and $\MsB$.
In (a), we show the regions where the {\it minimum} values of $R$,
obtained when scanning the other parameters, fulfill $R\geq 10^{-3}$.
In (b), we show the regions where the {\it maximum} values of $R$,
obtained when scanning the other parameters, fulfill $R\geq 10^{-3}$.
The region where $\MsB<50~\mbox{GeV}$ is not allowed.}

%%%%%%%%%%%%%%%%%%%%%%%%%%%%%%%%%%%%%%%%%%%%%%%%%%%%%%%%%%%%%%%%%%%%%%%%
%% possible shortcut:
\end{document}